\documentclass[prl,a4paper,twocolumn,showpacs,superscriptaddress,floatfix]{revtex4}
\usepackage{amsmath}
\usepackage{amsfonts}
\usepackage{amssymb}
\usepackage{bm}
\usepackage{latexsym}
\usepackage[dvips]{graphicx,color}

\newcommand{\rvec}{\mbox{$\underline{r}$}}
\newcommand{\uvec}{\mbox{$\underline{u}$}}

\newcommand{\duvec}{\mbox{$\underline{\delta u}$}}

\begin{document}

\title{Inhomogeneous elastic response of silica glass}

\author{F.~L\'eonforte}
\affiliation{Universit\'e Lyon I Laboratoire de Physique de la
Mati\`ere Condens\'ee et des Nanostructures;  CNRS, UMR 5586, 43
Bvd. du 11 Nov. 1918, 69622 Villeurbanne Cedex, France}
\author{A.~Tanguy}
\affiliation{Universit\'e Lyon I Laboratoire de Physique de la
Mati\`ere Condens\'ee et des Nanostructures; CNRS, UMR 5586, 43
Bvd. du 11 Nov. 1918, 69622 Villeurbanne Cedex, France}
\author{J.P.~Wittmer}
\affiliation{Institut Charles Sadron, CNRS, 6, Rue Boussingault,
67083 Strasbourg, France}
\author{J.-L.~Barrat$^1$}

\date{\today}

\begin{abstract}
Using large scale molecular dynamics simulations we investigate the
properties of the {\em non-affine} displacement field
induced by macroscopic uniaxial deformation of amorphous silica,
a strong glass according to Angell's classification.
We demonstrate the existence of a length scale $\xi$ characterizing the
correlations of this field (corresponding to a volume of about 1000 atoms),
and compare its structure to the one observed in a standard fragile model glass.
The ``Boson-peak'' anomaly of the density of states can be traced back
in both cases to elastic inhomogeneities on wavelengths smaller than $\xi$
where classical continuum elasticity becomes simply unapplicable.
\end{abstract}


\maketitle

\begin{figure}[t]
  \includegraphics*[width=8.5cm]{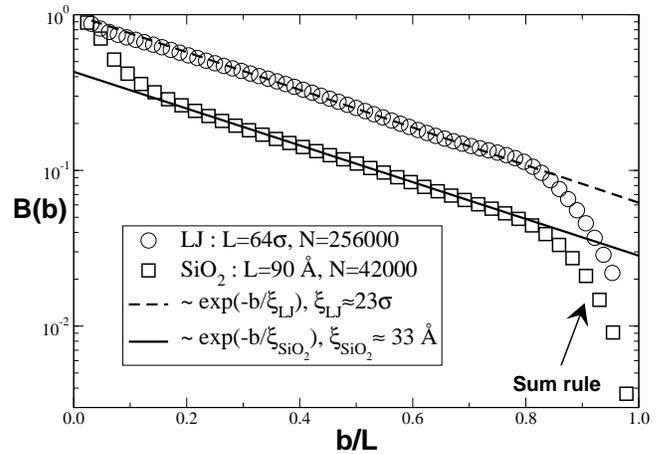}
  \vspace*{-0.2cm}
  \caption{\label{Bb_FIG} Amplitude of the coarse-graining
    function $B(b)$ of the normalized non-affine field averaged over a
    volume element of lateral size $b$, versus the ratio $b/L$, for LJ
    and silica (squares) systems under uniaxial elongation.
    Since the total displacement is zero by
    symmetry $B(b)$ must necessarily vanish for large $b \approx L$
    (``sum rule").
    For sufficiently large system sizes, allowing to probe a broad
    $\sigma/L \ll b/L \ll 1$ region, our data demonstrates an exponential
    decay with characteristic length scale $\xi\approx 23\sigma$ for LJ and
    $\xi\approx 33$ \AA ~for silica glasses.
   }
\end{figure}

\begin{figure}[t]
\includegraphics*[width=8.5cm]{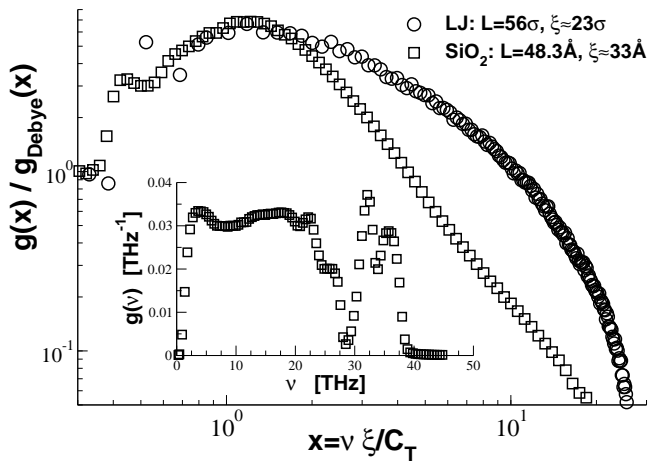}
\vspace*{-0.2cm}
\caption{\label{DOS_FIG}
Inset: VDOS $g(\nu)$ for the silica glass at $T=300$ K.
Main figure:
Excess of vibrational states $g(x)$ compared to
Debye's continuum model $g_{Debye}(x)$, using reduced
units $x=\nu \xi/C_T$, with $C_T\approx 4.2$ (in LJ units) and $C_T\approx 3961.4$m/s
for LJ and silica respectively and $\xi$ as indicated in the figure.
The Boson Peak position at $x \approx 1$ is well approximated by the
frequency associated with the wavelength of order $\xi$.
As expected, the peak is more pronounced for the strong glass.
}
\end{figure}

%
The vibrational dynamics of glasses and in particular
the vibrational anomaly known as the ``Boson Peak",
i.e. an excess of the low-energy density of state
in glasses relative to the Debye model,
have attracted considerable attention
in condensed matter physics \cite{Model_BP,Montpellier,Rome}.
This anomaly is observed in Raman and Brillouin
spectroscopy \cite{Raman_Brillouin} and inelastic neutron
scattering \cite{Neutron} experiments in many different
systems (polymer glasses \cite{Polymer}, silica \cite{Silica},
metallic glasses \cite{Metallic}) and the
corresponding excitations are often associated with heat capacity
or heat conductivity low temperature anomalies. Many
interpretations of these vibrational anomalies have been put forward,
and generally involve some kind of disorder generated
inhomogeneous behavior \cite{Model_BP}, whose exact nature,
however, is the subject of a lively
debate \cite{Montpellier,Rome,Raman_Brillouin,DynamicLength}.\\
%
%
\indent
In this work, we argue that the natural origin of these anomalies in
``fragile'' as well as ``strong'' glasses lies in the inhomogeneities of the
elastic response at small scales, which can be characterized through the
correlation length $\xi$ of the inhomogeneous or ``{\em non-affine}"
part of the displacement field generated in response to an
elastic deformation imposed at the macroscopic scale.
The existence of such a length has been suggested in a series of
previous numerical studies \cite{Wittmer,Leonforte_2,footLJ}
on two and three dimensional Lennard-Jones (LJ) systems, and is experimentally
demonstrated in macroscopic amorphous solids (foams \cite{Debre}, emulsions \cite{Falk},
granulars \cite{Goldenberg},
\ldots).
At a more microscopic level,  evidence has been provided recently by
UV Brillouin scattering experiments on amorphous silica \cite{Masciovecchio}.
Being a natural consequence of the disorder of microscopic interactions
\cite{Leonforte_2} the non-affine displacement field is responsible in
particular for the breakdown of Born-Huang's formulation \cite{Born_Huang}
for the prediction of elastic moduli \cite{Wittmer,Lemaitre,Lutsko94},
and has recently been studied theoretically by Lema\^itre {\em et al.}
\cite{Lemaitre} and DiDonna {\em et al.} \cite{DiDonna}.\\
%
%
%
\indent
In practice, however, it appears that the only practical way to quantify
this effect for a given material consists in direct molecular simulations
\cite{Wittmer,Leonforte_2}.
The present contribution extends, for the first time, the
numerical analysis to a realistic model of an amorphous silca melt
--- a ``strong" glass according to Angell's classification
\cite{Angell}. Our results are compared to a previously studied
``fragile'' reference glass formed by weakly polydisperse LJ
particles in 3D \cite{footLJ}. Strong and fragile systems have
very different molecular organisation and bonding. Although the
intensity of vibrational anomalies is less important in fragile
systems, it is well documented in experiments \cite{Polymer} on
polymer glasses or in simulations of Lennard-Jones systems \cite{glass_general}.
The observation of common features points to a universal framework for
the description of low frequency vibrations in glassy systems.
One recent finding of particular interest, is the fact that, in
these LJ systems, the Boson Peak anomaly appears to be located at
the {\em edge} of the non-affine displacement regime, its position
given by the pulsation associated with $\xi$ \cite{Leonforte_2}.
This begs the question whether this is a {\em generic} result
applying also to other glasses, specifically to strong glass
forming materials such as amorphous silica, which is characterized
by an intricate local packing \cite{glass_general} --- believed
widely to be the {\em specific} origin of the vibrational anomaly
\cite{Montpellier}.
As in our earlier contributions, we will focus on the analysis of the
non-affine displacement field obtained in the {\em linear elastic} strain
regime and the eigenmode density of states for systems at zero temperature
or well below the glass transition.\\
%
\indent
The amorphous silica is modelled using the force field proposed by
van Beest {\em et al.} \cite{BKS}. (For details about this ``BKS" potential,
see Refs.~\cite{Vollmayr}.)
We performed classical NVT Molecular Dynamics simulations of systems containing
$N=8016,\,24048$ and $42000$ atoms with density
$\rho=2.37\textrm{ g/cm}^3$, in  fully periodic cubic cells with
sizes $L=48.3\textrm{ \AA},\,69.6\textrm{ \AA}$ and
$83.8\textrm{\AA}$, respectively.
%
The short ranged part of the BKS potential was truncated and
shifted at a distance of $5.5$ \AA.
For the Coulomb part we use the Ewald method with a real-space sum
truncated and shifted at $9$ \AA~ \cite{Plimpton}.
To obtain the  silica glass, we first
equilibrate all systems at $T=5000$ K during $0.8$ ns. An ensemble
of three independent configurations was studied for each system size
\cite{footselfaverage}.
Next, we perform a quench from $T=5000$ K to $T=0$ K by
decreasing linearly the temperature of the external heat bath with
a quench rate of $1.8$ K/ps \cite{Vollmayr}. Finally, a
Conjugate Gradient algorithm was used to minimize the potential
energy of the systems yielding $T=0$ K configurations with
hydrostatic pressure $\langle P\rangle\approx 0.4$ GPa.
The static properties were checked against published results
obtained with the same amorphous silica model \cite{Vollmayr}.\\
%
%
\begin{figure}[t]
\includegraphics*[width=6.5cm]{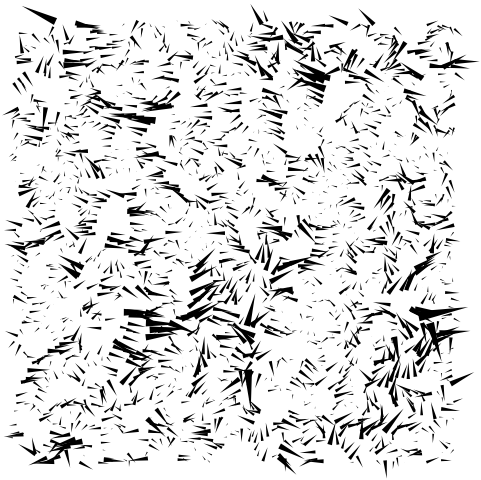}
\caption{\label{NAFF_FIG}
Inhomogeneous part $\duvec(\rvec)$ of
the displacement field $\uvec(\rvec)$ for
the imposed macroscopic uniaxial strain in elongation
$\epsilon_{xx}=5.10^{-3}$ for a silica glass containing $N=42000$
particles ($L=83.8$ \AA). Projection of the field is done on
the $(x-z)$-plane for all particles with position $\rvec$ close to the plane,
the arrow length being proportional to $\duvec(\rvec)$.
The field resides in the linear elastic regime, i.e. has a magnitude
varying linearly and reversibly with the applied deformation.
As visual inspection shows, it is strongly spatially correlated and
involves a substantial fraction of all atoms.
}
\end{figure}
%
%
%
\indent
We now describe briefly the protocol used in order to investigate the
elastic behavior at zero temperature of the model glasses under
uniaxial deformation
(for more details, see Refs.~\cite{Wittmer,Leonforte_2}).
The procedure consists in applying a global deformation of strain
$||\epsilon_{xx}||\ll 1$ to the sample by rescaling all coordinates in
an {\em affine} manner. Starting from this affinely deformed configuration,
the system is relaxed to the nearest energy minimum, keeping the shape of
the simulation box constant. The relaxation step releases about
half of the elastic energy of the initial affine deformation and results
in the displacement $\duvec(\rvec)$ of the atoms relative to the affinely
deformed state, defining the non-affine displacement field.
A typical field for a silica glass is presented in Fig.~\ref{NAFF_FIG},
where a 2D projection of $\duvec(\rvec)$ in the plane containing the applied
deformation direction is shown.
\\
%
%
\indent
This procedure allows us to measure directly the elastic coefficients
from Hooke's law \cite{Wittmer}, i.e. from the stress differences
$\Delta\sigma_{\alpha\beta}=\sigma^{end}_{\alpha\beta}-\sigma^{ref}_{\alpha\beta}$,
$\sigma^{ref}_{\alpha\beta}$ being the total stress tensor of the
reference state configuration (quenched stresses), and
$\sigma^{end}_{\alpha\beta}$ the one measured in the deformed
configuration after relaxation. From the resulting values of the
Lam\'e coefficients $\lambda=\Delta\sigma_{yy}/\epsilon_{xx}$ and
$\mu=(\Delta\sigma_{xx}-\Delta\sigma_{yy})/2\epsilon_{xx}$
one obtains the associated transverse and
longitudinal sound wave velocities,
$C_T=\sqrt{\mu/\rho}$, $C_L=\sqrt{(\lambda+2\mu)/\rho}$.
In the case of the silica glass
($\lambda \approx 34.4$ GPa,
$\mu\approx 37.2$ GPa,
$C_T \approx 3961.4$ m/s,
$C_L \approx 6774.5$ m/s),
these quantities are in good agreement with data from Horbach {\em
et al.} \cite{Vollmayr} and Zha {\em et al.} \cite{Zha} (for silica
under a density of $2.2\textrm{g/cm}^3$, taking into account the
scaling factor
$(2.37/2.2)^{1/2}$ \cite{Vollmayr} inherent to the choice of a higher density).\\
%
%
\indent The {\em linearity} of the strain dependence of both the
displacement field and the stress difference
$\Delta\sigma_{\alpha\alpha}$ have been verified explicitly
following Ref.~\cite{Leonforte_2}. The {\em elastic} (reversible)
character of the applied deformation is checked by computing the
remaining residual displacement field after removing the external
strain \cite{Leonforte_2}.
An alternative quantification of the plastic deformation is obtained by
considering the participation  ratio $Pr=N^{-1}\left(\sum_i
\duvec^2_i\right)^2/\sum_i \left(\duvec^2_i\right)^2$ of the noise
$\duvec(\rvec)$ \cite{Leonforte_2}. As long as  $Pr\approx 1$, all
atoms are involved in the non-affine field, while
irreversible plastic rearrangements are marked by $Pr\rightarrow 0$,
with only a few particles involved. A choice of
$\epsilon_{xx}=10^{-7}$ for the LJ glass with $L=56\sigma$ and of
$\epsilon_{xx}=5.10^{-3}$ for the silica glass were found to ensure
\textit{reversible and linear } behavior, with $20\%<Pr<30\%$ and
$25\%<Pr<40\%$, respectively \cite{footlinelastic}.
\\
%
%
\begin{figure}[t]
\includegraphics*[width=8.5cm]{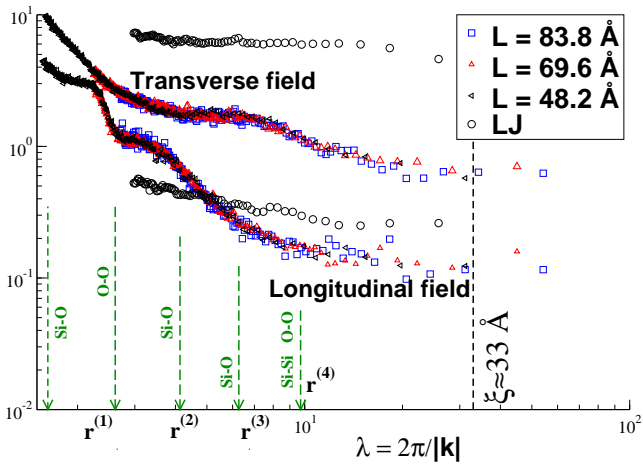}
\vspace*{-0.2cm}
\caption{\label{STLK_FIG} (Color online) Squared amplitudes
  of the Fourier transforms $S_L(k)$ and $S_T(k)$ for the
  longitudinal (bottom) and transverse contributions to the
  {\em normalized} non-affine field $\duvec(\rvec)$ of silica glass
  at $T=0$ K under macroscopic elongation, plotted {\em vs.}
  the wavelength $\lambda=2\pi/k$ (in \AA). The various system sizes
  included demonstrate a prefect data collapse.
  The transverse contribution is more important for all wavelengths.
  The spectra become constant for large wavelengths with a
  relative amplitude of about 10.
  The spectra for a LJ glass (spheres, length given in sphere diameters)
   have been included for comparison \cite{footSTLKconst}
}
\end{figure}
\indent
Visual inspection of the snapshot suggests that the field is strongly
correlated over large distances, with the presence of rotational structures
previously observed in Ref.~\cite{Leonforte_2} for LJ systems \cite{footnonaff}.
In order to characterize this kind of structure, we
normalize the 
field by its second moment, i.e.
$\duvec(\rvec)\mapsto\duvec(\rvec)/\langle\duvec(\rvec)^2\rangle^{1/2}$.
In this way, in the linear elastic regime,
it becomes independent of the applied strain and the system size \cite{DiDonna}.
\\
\indent
Next, we study the Fourier power spectrum of the fluctuations of this
normalized field. This spectrum can be described by
two structure factors,
$S_L(k)\equiv \langle\parallel\sum_{j=1}^N
\hat{k} \cdot\duvec(\rvec_j)\exp{(i\underline{k}\cdot
\rvec_j)}\parallel^2\rangle / N$ relative to the longitudinal and
$S_T(k)\equiv \langle\parallel\sum_{j=1}^N
\hat{k} \wedge\duvec(\rvec_j)\exp{(i\underline{k}\cdot
\rvec_j)}\parallel^2\rangle / N$  relative to the transverse
field component \cite{Leonforte_2}.
These quantities are plotted in
Fig.~\ref{STLK_FIG} as function of the wavelength
$\lambda=2\pi/k$, where $\underline{k}= k \ \hat{k}=(2\pi/L)(l,m,n)$
with $\hat{k}$ being the normalized wavevector.
Brackets
$\langle\cdot\rangle$  denote an average  over the degeneracy set
associated with $\lambda$, and over an ensemble of configurations.
As expected from our study of LJ glasses \cite{Leonforte_2},
the longitudinal power spectrum of silica is always smaller than the transverse one.
The main difference between the two materials resides in the
{\em hierarchical progression} of the decoupling between
transverse and longitudinal contributions at short wavelengths that
appears in the case of the silica glass
(the spectra of LJ systems being only weakly wavelength dependent).
This can be traced back to the local structure of silica which is
represented  by arrows giving the positions of the $n$ first neighbor shells
$r^{(n)}_{\{\alpha-\beta\}}$, where $n\in[1,4]$ and $(\alpha,\beta)\in\{Si,O\}$.
Structural effects disappear at distances greater than 4-5 tetrahedral
units $SiO_4$, i.e. $r^{(4-5)}_{\{\alpha-\alpha\}}$ with $\alpha\in\{Si,O\}$,
and the longitudinal contribution to the non-affine displacement
field becomes then about 10 times smaller than the transverse one ---
similar to our finding for LJ systems \cite{Leonforte_2,footSTLKconst}.\\
%
%
%
%
\indent
We conclude that the non-affine displacement field is of predominantly
{\em rotational} nature in both ``fragile'' and ``strong'' glasses, and
proceed to extract a characteristic length representative of this rotational
structure. Considering the coarse-grained field
$\underline{U}_j(b)\equiv N^{-1}_j\sum_{i\in V_j}\duvec(\rvec_i)$
of all $N_j$ particles contained within a cubic volume element $V_j$
of linear size $b$, we compute the coarse-graining function
$B(b)\equiv\langle\underline{U}_j(b)^2\rangle_j^{1/2}$.
As shown by Fig.~\ref{Bb_FIG}, we find for {\em both} glasses
an exponential decay, well fitted by the characteristic
scales $\xi\approx 23\sigma$ for the ``fragile'' glass, and $\xi\approx 33$
\AA, i.e. near $23\times r^{(1)}_{\{Si-O\}}$ for the ``strong'' glass.
The latter length scale has also been indicated in Fig.~\ref{STLK_FIG}.
The exponential behavior becomes more pronounced with increasing system
size (not shown) which reduces the regime of the cut-off observed at large
$b/L \approx 1$ which is expected from the symmetry of the total non-affine field.
($B(b)\rightarrow 1$ for $b\rightarrow 0$ due to the normalization of the field.)
\\
%
%
%
\indent The existence of such a characteristic length scale has
already been underlined in Ref.~\cite{Leonforte_2} for the LJ
system, and has been related to the position of the Boson Peak in
the density of vibrational states. In order to test this
assumption in the case of the silica glass, we computed the
vibrational density of states (VDOS) $g(\nu)$ using the Fourier
transform of the velocity autocorrelation function \cite{Vollmayr},
calculated during 1.6 ns at $T=300$ K (followed after a run of 8
ns to assure equipartition of the kinetic energy at this
temperature). The VDOS  is shown in the inset of the
Fig.~\ref{DOS_FIG}, and is in good agreement with results from
\cite{Vollmayr}. In the main part of Fig.~\ref{DOS_FIG}, reduced
units $x=\nu\times \xi/C_T$ are used in order to plot the excess
of vibrational states according to Debye's continuum prediction,
i.e. $g(x)/g_{Debye}(x)$, with $\xi$ the previous characteristic
length scales and $C_T$ the sound velocities for transverse waves,
for LJ and silica glasses. (The Debye prediction must obviously
become correct for small eigenfrequencies. To access this
frequency range  even larger simulation boxes are needed.) This
plot confirms the fact that the Boson Peak position can be well
approximated by the frequency associated with the correlation
length $\xi$ of elastic heterogeneities in both LJ and silica
glasses.
\\
%
%
\indent
In summary, we have demonstrated the existence of inhomogeneous and
mainly rotational rearrangements in the elastic response to a macroscopic
deformation of amorphous silica. Our results are similar to the ones
obtained previously for LJ glasses.
The characterization of the non-affine displacement field demonstrates the
existence of correlated displacements of about 1000 particles corresponding
to elastic heterogeneities of characteristic size $\xi$ of 20 interatomic
distances.
The estimate of the frequency associated with this length is in good agreement
with the Boson Peak position.
The existence of such a characteristic length in glasses should encourage
to view the Boson Peak as a length --- rather than a frequency ---
marking the crossover between a regime where vibrations in glasses with
wavelengths larger than $\xi$ can be well described by a classical
continuum theory of elasticity, and a small wavelength regime where
vibrations are strongly affected by elastic heterogeneities.
\\
\indent
In a nutshell, the vibrational anomaly is therefore simply due to physics
on scales where classical continuum elastic theories (such as the Debye model)
must necessarily break down. This leaves unanswered the important question
what additional excitations are probed that produce the peak but suggests
a similar description for different glass formers.
Interestingly, the existence of a length scale of comparable magnitude
accompanying the glass transition of liquids as been demonstrated very
recently \cite{DynamicLength}. This (dynamical) length characterizes the
number of atoms which have to move simultaneously to allow flow just
as our (static) length $\xi$ describes the correlated particle displacements.
Since the glass structure is essentially frozen at the glass transition
both correlations may be closely related, possibly such that the
non-affine displacements might be shown in future work to be
reminiscent of the dynamical correlations at the glass transition.

\begin{acknowledgments}
Computer time was provided by IDRIS, CINES and FLCHP.
\end{acknowledgments}


\begin{thebibliography}{99}
\bibitem{Model_BP}
See the reference lists of:
V.L.~Gurevich {\em et al.}, Phys. Rev. B {\bf 67}, 094203 (2003);
J.~Horbach {\em et al.}, Eur. Phys. J. B {\bf 19}, 531 (2001).
%
\bibitem{Montpellier}
M.~Foret {\em et al.}, Phys. Rev. B {\bf 66}, 24204 (2002);
B.~Ruff\'e {\em et al.}, Phys. Rev. Lett. {\bf 90}, 095502 (2003);
E. Courtens {\em et al.}, J. Phys.: Condens. Matter {\bf 15}, 1279 (2003).
%
\bibitem{Rome}
G. Ruocco {\em et al.}, J. Phys.: Condens. Matter {\bf 13}, 9141 (2001);
O. Pilla {\em et al.}, Phys. Rev. Lett. {\bf 85}, 2136 (2000);
P. Benassi {\em et al.}, Phys. Rev. Lett. {\bf 78}, 4670 (1997).
%
\bibitem{Raman_Brillouin}
R.S.~Krishnan, Proc. Indian Acad. Sci. A {\bf 37}, 377 (1953);
M.~Yamaguchi, T.~Nakayama, T.~Yagi, Physica B {\bf 263}, 258-260 (1999);
B.~Helen {\em et al.}, Phys. Rev. Lett. {\bf 84}, 5355 (2000);
B. Ruffl\'e {\em et al.}, Phys. Rev. Lett. {\bf 96}, 045502 (2006).
%
\bibitem{Neutron}
U.~Buchenau {\em et al.}, Phys. Rev. Lett. {\bf 53}, 2316 (1984);
M.T.~Dove {\em et al.}, Phys. Rev. Lett. {\bf 78}, 1070 (1997);
E.~Duval {\em et al.}, J. Non-Cryst. Solids {\bf 235}, 203 (1997).
\bibitem{Polymer}
E.~Duval, {\em et al.}, Europhys. Lett. {\bf 63}, 778 (2003).
\bibitem{Silica}
E.~Duval, {\em et al.}, Phil. Mag. {\bf 84}, 1433 (2004).
\bibitem{Metallic}
M.~Arai, {\em et al.}, Phil. Mag. B {\bf 79}, 1733 (1999).
\bibitem{DynamicLength}
L. Berthier, {\em et al.}, Science {\bf 16}, 1797 (2005).
\bibitem{Wittmer}
J.P.~Wittmer {\em et al.}, Europhys. Lett. {\bf 57}, 423 (2002);
A.~Tanguy {\em et al.}, Phys. Rev. B {\bf 66}, 174205 (2002).
\bibitem{Leonforte_2}
F.~L\'eonforte {\em et al.}, Phys. Rev. B, {\bf 72}, 224206 (2005).
\bibitem{footLJ}
A LJ pair potential
$U_{ij}(r)=4\epsilon\left((\sigma_{ij}/r)^{12}-(\sigma_{ij}/r)^6\right)$
for slightly polydisperse beads is used with $\sigma_{ij}$ uniformly
distributed between $0.8\sigma$ and $1.2\sigma$. All LJ data presented
in this paper refer to at a fixed density $\rho\sigma^3=0.98$ corresponding
to a hydrostatic pressure $\langle P\rangle\approx 0.2$ at zero temperature.
See Ref.~\cite{Leonforte_2} for details concerning the simulation protocol.
%
\bibitem{Debre}
G.~Debr\'egeas, H.~Tabuteau and J.-M. diMeglio, Phys. Rev. Lett. {\bf 87}, 178305 (2001).
\bibitem{Falk}
M. Falk, J. Langer, Phys. Rev. E {\bf 57}, 7192 (1998).
\bibitem{Goldenberg}
C.~Goldenberg {\em et al.}, preprint cond-mat/0511610.
%
\bibitem{Masciovecchio}
G.~Masciovecchio {\em et al.},
``Evidence for a new nano-scale stress correlation in silica'', preprint 2005.
\bibitem{Born_Huang}
K.~Huang, Proc. Roy. Soc. London A {\bf 203}, 178 (1950);
M.~Born and K.~Huang,
{\em Dynamical Theory of Crystal Lattices} (Clarendon Press, Oxford, 1954).
\bibitem{Lutsko94}
J.F.~Lutsko, J. Appl. Phys. {\bf 65}, 2991 (1989).
\bibitem{Lemaitre}
A.~Lema\^itre and C.~Maloney, cond-mat/0410592.
\bibitem{DiDonna}
B.A.~DiDonna and T.C.~Lubensky, Phys. Rev. E {\bf 72},  066619,
(2005).
\bibitem{Angell}
C.A.~Angell, Science {\bf 267}, 1924 (1995).
\bibitem{glass_general}
W. Kob, K. Binder, {\em Glassy materials and disordered solids}
(World Scientific Publishing, 2005).
\bibitem{BKS}
B.W.H.~van Beest, G.J.~Kramer, and R.A.~van Santen,
Phys. Rev. Lett. {\bf 64}, 1955 (1990).
\bibitem{Vollmayr}
K.~Vollmayr, W.~Kob, and K.~Binder, Phys. Rev. B {\bf 54}, 15808 (1996);
J.~Horbach, W.~Kob, and K.~Binder, J. Phys. Chem. B {\bf 103}, 4104 (1999).
\bibitem{Plimpton}
We use the implementation in the LAMMPS code described in
S.J.~Plimpton, J. Comp. Phys. {\bf 117}, 1 (1995).
\bibitem{Zha}
C.-S.~Zha {\em et al.}, Phys. Rev. B {\bf 50}, 13105 (1994).
%
\bibitem{footselfaverage}
The sample to sample variations of all properties discussed in this study
were found to be small. This is expected since all quantities are
self-averaging and $\xi \approx 33$ \AA $\ll L$.
%
\bibitem{footnonaff}
The affine part of the displacement field,
$\underline{\underline{\epsilon}}\cdot \rvec$, depends trivially on the
bead position within the simulation box. As may be seen from Fig.~\ref{NAFF_FIG},
the non-affine part does {\em not}.
%
\bibitem{footlinelastic}
The linear elastic regime for amorphous silica is much broader
(higher plasticity threshold).
%
\bibitem{footSTLKconst}
For both model glasses the Fourier spectra  become constant for
large wavelengths. This is similar to the structure factor
associated with the density fluctuations and reflects the thermal
fluctuations  which are frozen in at $T_g$. The  long wavelength
limit should be of order $\rho \xi^3$ where $\rho$ is the number
density and $\xi$ the correlation length of the field. This
explains the difference in magnitude between the two systems.
\end{thebibliography}
\end{document}